\documentclass[aps,prl,twocolumn,showpacs,superscriptaddress,groupauthors,nobibnotes]{revtex4-1}  
\usepackage{graphicx}  
 
\begin{document}
\widetext

\title{Transverse Single-Spin Asymmetry and Cross-Section for $\pi^0$ and $\eta$ Mesons at Large Feynman-$x$ in $p^{\uparrow}+p$ Collisions at $\sqrt{s}=200$ GeV}

\affiliation{AGH University of Science and Technology, Cracow, Poland}
\affiliation{Argonne National Laboratory, Argonne, Illinois 60439, USA}
\affiliation{Brookhaven National Laboratory, Upton, New York 11973, USA}
\affiliation{University of California, Berkeley, California 94720, USA}
\affiliation{University of California, Davis, California 95616, USA}
\affiliation{University of California, Los Angeles, California 90095, USA}
\affiliation{Universidade Estadual de Campinas, Sao Paulo, Brazil}
\affiliation{Central China Normal University (HZNU), Wuhan 430079, China}
\affiliation{University of Illinois at Chicago, Chicago, Illinois 60607, USA}
\affiliation{Cracow University of Technology, Cracow, Poland}
\affiliation{Creighton University, Omaha, Nebraska 68178, USA}
\affiliation{Czech Technical University in Prague, FNSPE, Prague, 115 19, Czech Republic}
\affiliation{Nuclear Physics Institute AS CR, 250 68 \v{R}e\v{z}/Prague, Czech Republic}
\affiliation{University of Frankfurt, Frankfurt, Germany}
\affiliation{Institute of Physics, Bhubaneswar 751005, India}
\affiliation{Indian Institute of Technology, Mumbai, India}
\affiliation{Indiana University, Bloomington, Indiana 47408, USA}
\affiliation{Alikhanov Institute for Theoretical and Experimental Physics, Moscow, Russia}
\affiliation{University of Jammu, Jammu 180001, India}
\affiliation{Joint Institute for Nuclear Research, Dubna, 141 980, Russia}
\affiliation{Kent State University, Kent, Ohio 44242, USA}
\affiliation{University of Kentucky, Lexington, Kentucky, 40506-0055, USA}
\affiliation{Institute of Modern Physics, Lanzhou, China}
\affiliation{Lawrence Berkeley National Laboratory, Berkeley, California 94720, USA}
\affiliation{Massachusetts Institute of Technology, Cambridge, MA 02139-4307, USA}
\affiliation{Max-Planck-Institut f\"ur Physik, Munich, Germany}
\affiliation{Michigan State University, East Lansing, Michigan 48824, USA}
\affiliation{Moscow Engineering Physics Institute, Moscow Russia}
\affiliation{Ohio State University, Columbus, Ohio 43210, USA}
\affiliation{Old Dominion University, Norfolk, VA, 23529, USA}
\affiliation{Panjab University, Chandigarh 160014, India}
\affiliation{Institute of Nuclear Physics PAS, Cracow, Poland}
\affiliation{Pennsylvania State University, University Park, Pennsylvania 16802, USA}
\affiliation{Institute of High Energy Physics, Protvino, Russia}
\affiliation{Purdue University, West Lafayette, Indiana 47907, USA}
\affiliation{Pusan National University, Pusan, Republic of Korea}
\affiliation{University of Rajasthan, Jaipur 302004, India}
\affiliation{Rice University, Houston, Texas 77251, USA}
\affiliation{Universidade de Sao Paulo, Sao Paulo, Brazil}
\affiliation{University of Science \& Technology of China, Hefei 230026, China}
\affiliation{Shandong University, Jinan, Shandong 250100, China}
\affiliation{Shanghai Institute of Applied Physics, Shanghai 201800, China}
\affiliation{SUBATECH, Nantes, France}
\affiliation{Texas A\&M University, College Station, Texas 77843, USA}
\affiliation{University of Texas, Austin, Texas 78712, USA}
\affiliation{University of Houston, Houston, TX, 77204, USA}
\affiliation{Tsinghua University, Beijing 100084, China}
\affiliation{United States Naval Academy, Annapolis, MD 21402, USA}
\affiliation{Valparaiso University, Valparaiso, Indiana 46383, USA}
\affiliation{Variable Energy Cyclotron Centre, Kolkata 700064, India}
\affiliation{Warsaw University of Technology, Warsaw, Poland}
\affiliation{University of Washington, Seattle, Washington 98195, USA}
\affiliation{Wayne State University, Detroit, Michigan 48201, USA}
\affiliation{Yale University, New Haven, Connecticut 06520, USA}
\affiliation{University of Zagreb, Zagreb, HR-10002, Croatia}

\author{L.~Adamczyk}\affiliation{AGH University of Science and Technology, Cracow, Poland}
\author{G.~Agakishiev}\affiliation{Joint Institute for Nuclear Research, Dubna, 141 980, Russia}
\author{M.~M.~Aggarwal}\affiliation{Panjab University, Chandigarh 160014, India}
\author{Z.~Ahammed}\affiliation{Variable Energy Cyclotron Centre, Kolkata 700064, India}
\author{A.~V.~Alakhverdyants}\affiliation{Joint Institute for Nuclear Research, Dubna, 141 980, Russia}
\author{I.~Alekseev}\affiliation{Alikhanov Institute for Theoretical and Experimental Physics, Moscow, Russia}
\author{J.~Alford}\affiliation{Kent State University, Kent, Ohio 44242, USA}
\author{B.~D.~Anderson}\affiliation{Kent State University, Kent, Ohio 44242, USA}
\author{C.~D.~Anson}\affiliation{Ohio State University, Columbus, Ohio 43210, USA}
\author{D.~Arkhipkin}\affiliation{Brookhaven National Laboratory, Upton, New York 11973, USA}
% ...

\collaboration{STAR Collaboration }\noaffiliation

\date{\today}

\begin{abstract}
Measurements of the differential cross-section and the transverse single-spin asymmetry, $A_{N}$, vs. $x_F$ for $\pi^0$ and $\eta$ mesons are reported for $0.4 < x_F < 0.75$ at an average pseudorapidity of 3.68. A data sample of approximately 6.3 pb$^{-1}$ was analyzed, which was recorded during $p^{\uparrow}+p$ collisions at $\sqrt{s}$ = 200 GeV by the STAR experiment at RHIC. The average transverse beam polarization was 56\%. The cross-section for $\pi^0$ is consistent with a perturbative QCD prediction, and the $\eta/\pi^0$ cross-section ratio agrees with previous mid-rapidity measurements. For $0.55 < x_F < 0.75$, $A_{N}$ for $\eta$ ($0.210 \pm 0.056$) is 2.2 standard deviations larger than $A_N$ for $\pi^0$ ($0.081 \pm 0.016$). 
\end{abstract}

\maketitle

A well known prediction of collinearly factorized perturbative Quantum Chromodynamics (pQCD) is that the cross-section for forward meson production in proton-proton collisions should have negligible dependence on the transverse polarization of the incident proton \cite{Kane:1978nd}. This early prediction was contradicted by measurements \cite{ZGS:1976,AGS:1990,Adams:1991rw,Adams:1991cs,Allgower:2002qi} of sizable pion transverse single-spin asymmetries ($A_N$), defined for a forward moving polarized beam scattering to the left and with a vertical spin quantization axis as
\begin{equation}
A_N\equiv \frac{\sigma^\uparrow-\sigma^\downarrow}{\sigma^\uparrow+\sigma^\downarrow}.
\label{eq:AN_def}
\end{equation}

In order to explain the large asymmetries, several extensions of the pQCD collinear framework have been proposed. These approaches take into account the possible spin-dependent transverse components of parton momentum (Sivers effect \cite{Sivers:1989cc}), the possible spin-dependent fragmentation of a scattered polarized parton (Collins effect \cite{Collins:1992kk}), or higher-twist effects where transverse momenta related to the previous approaches are included in the hard scattering term of a collinear calculation \cite{Qiu:1991pp,Kouvaris:2006zy,Kang:2011hk}. A wide range of high energy polarized experiments, both nucleon-nucleon \cite{:2008qb,Adler:2005in,:2008mi} and lepton-nucleon \cite{Airapetian:2004tw,:2008dn,Airapetian:2009hr,Alekseev:2010cp}, have been performed to characterize the kinematic and process dependences of the asymmetries, and in the case of the latter, to directly test these approaches. 

For more than 20 years, we have known that the transverse asymmetries in forward pion production depend critically on the isospin projection ($I_3$) of the produced mesons relative to that of the parent hadron. In proton scattering experiments, the asymmetry for the $\pi^-$ meson, which contains a down quark and an anti-up quark, has the opposite sign relative to the asymmetries for the $\pi^+$ and $\pi^0$ mesons, produced from the predominant up quarks. 

In this paper, we report for the first time at $\sqrt{s}$ = 200 GeV the transverse single-spin asymmetry for the $\eta$ meson, another member of the pseudo-scalar octet that has the same isospin projection as the $\pi^0$ ($I_3=0$). We note that the FNAL-E704 collaboration previously found a large $A_N$ for the $\eta$ for Feynman-$x$ (longitudinal momentum of the observed particle divided by the beam energy) $ > 0.4$ at $\sqrt{s}$ = 19.4 GeV \cite{Adams:1997dp}. In addition, we report the differential cross-section for $\eta$ production in the region where the spin asymmetry is measured.

Leading-twist collinear pQCD has been successful in describing a wide range of unpolarized cross-section measurements at the Relativistic Heavy Ion Collider (RHIC), from $\pi^0$ at forward rapidity \cite{Adams:2003fx, Adams:2006uz}, to $\pi^0$ \cite{Adare:2007pi, Adare:2008pi, Abelev:2009pi, Abelev:2010pi}, $\eta$ \cite{Adler:2006bv, Abelev:2010pi, Adare:2011eta}, $\pi^\pm / K^\pm$ \cite{Arsene:2007jd}, and jets \cite{PhysRevLett.97.252001} at mid-rapidity. Such agreement is considered a strong indicator that the given process can be interpreted within the framework of pQCD. Therefore, the comparison between the unpolarized cross-section and the leading-twist collinear pQCD prediction becomes the basis on which to apply the aforementioned theoretical extensions to the associated transverse spin effects.

For forward $\pi^0$ production, recent STAR measurements of the cross-section are consistent with next-to-leading-order (NLO) pQCD calculations in the same region where a large transverse spin asymmetry is found \cite{Adams:2003fx, Adams:2006uz}. However, these results do not cover the large Feynman-$x$ ($x_F$) region where the acceptance for the $\eta$ decaying into two photons becomes large. In this paper, we have extended the analysis of the $\pi^0$ cross-section and $A_N$ to $x_F$ of 0.75, where its spin asymmetry and cross-section can be directly compared to those of the $\eta$ mesons.

The data were taken with the STAR Forward Pion Detector (FPD). The FPD is a modular lead glass calorimeter located at forward rapidity in the STAR interaction region at RHIC at Brookhaven National Laboratory. Two modules were placed on either side of the beam line, covering the pseudorapidity region from approximately 3.3 to 4.0. Each module contained 49 cells (glass blocks approximately 18 radiation lengths deep), forming a $7 \times 7$ square array. The data were collected in 2006 with transversely polarized proton beams and an integrated luminosity of $\sim 6.3$ pb$^{-1}$. The average polarization was (56.0\,$\pm$\,2.6)\% for the beam facing the FPD. As $A_N$ is a single spin observable, the spin state of the second beam (with (55.0\,$\pm$\,2.6)\% polarization) was integrated over for the $A_N$ at positive $x_F$. At negative $x_F$, the spin state of the first beam was integrated over. Events were recorded only when the total ADC count in either of the two modules was greater than a fixed threshold, which was nominally equivalent to 30 GeV. Photons reconstructed within a quarter of a cell from the detector edge were discarded. Only those events with exactly two reconstructed photons were analyzed, with the resulting loss of yield corrected for the cross-section measurement. The STAR Beam Beam Counter (BBC) on the away side was used to reject the single-beam background. The near side BBC was not required to produce a signal, as most of the analyzed events already had more than half the beam energy deposited in the FPD. The efficiency for the away side BBC condition was estimated to be (93\,$\pm$\,4)\% for all non-singly diffractive events based on previous analyses \cite{PhysRevLett.91.172302}.

\begin{figure}[t]
\includegraphics[width=.48\textwidth]{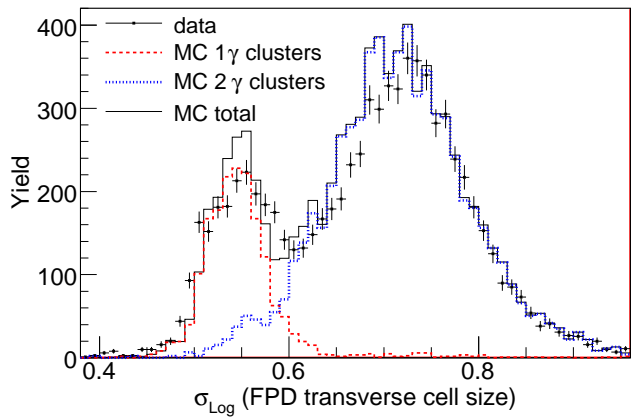}
\caption{(color online) The distribution of $\sigma_{Log}$, as defined in the text, for $E_{cluster} > 65$ GeV for data and simulation, in units of FPD transverse cell size (3.81 cm). For comparison, the one and two photon cluster peaks from simulation were independently normalized, and uniformly shifted by $+0.01$ transverse cell size (1 bin) to account for the small difference in the average size of clusters between simulation and data.}
\label{sigma} 
\end{figure}

The $x_F$ coverage of the previous analysis of $A_N$($\pi^0$) \cite{:2008qb}, which included this data set, was limited by the difficulty in separating $\pi^0$ clusters from single photons with $x_F > 0.55$. At this point, the typical separation of two $\pi^0$ decay photons at the surface of the FPD becomes similar to the Moli\`{e}re radius of the lead glass (3.32 cm) and transverse cell size (3.81 cm). On the other hand, the $\eta$ meson acceptance lies mostly above an $x_F$ of 0.5 due to the larger separation of its decay photons. 

With the current analysis, the $\pi^0$ - $\gamma$ separation has been greatly improved by analyzing $\sigma_{Log}$, defined as
\begin{equation}
\sigma_{Log} \equiv \sqrt{ \frac{\sum_{i} Log[(E_i+E_0)/GeV] \cdot (\bar{x}-x_i)^2} {\sum_{i} Log[(E_i+E_0)/GeV]} } ,
\label{eq:sigma_def}
\end{equation}
where $E_i$ and $x_i$ are the energy and the location of the $i^{th}$ channel, and $E_0$ = 0.5 GeV. The $i^{th}$ term in the sum is skipped if $Log(E_i + E_0) < 0$. It provides a significant sensitivity to the topological differences between single and double photon clusters at high energies, as evidenced by the clear separation between the one and two photon peaks in Fig. \ref{sigma}. Also shown are the results from \textsc{pythia} and \textsc{geant} simulations, which closely reproduce the $\sigma_{Log}$ distributions for both types of clusters up to a small offset ($\sim 1$\% of the transverse cell size). As a result, the $x_F$ coverage for $\pi^0$'s was extended from 0.55 to 0.75.

In addition, the \textsc{geant} simulation of the electromagnetic shower in the FPD is now based on the tracking of optical photons produced by the Cherenkov effect. Compared to the previous method based on charged particle energy loss, the new simulation produces a better agreement with the data on shower shape, energy resolution, and the observed shift in gain as a function of photon energy. Combined with a more advanced parameterization of the shower shape including the effects of incident angle, it allows for a higher precision calibration needed for the cross-section measurement.  

\begin{figure}[t]
\includegraphics[width=.48\textwidth]{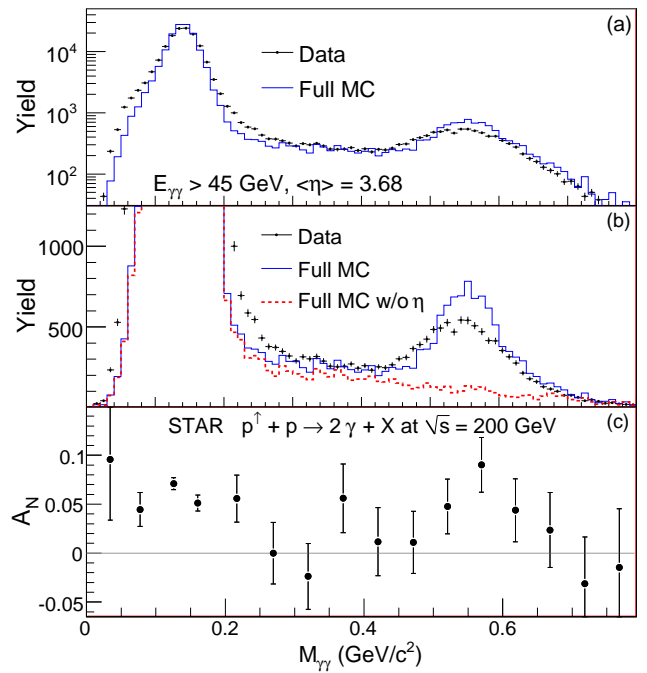}
\caption{(color online) (a) Di-photon invariant mass, $M_{\gamma\gamma}$, distributions in data and simulation for $E_{\gamma\gamma} > 45$ GeV, with the ``center cut" as defined in Eq.\@ (\ref{eq:center}). The simulation results were normalized to have the same number of events as the data in the $\pi^0$ mass region ($0.08 < M_{\gamma\gamma} < 0.19$ GeV/$c^2$).  The symbol $\langle \eta \rangle$ indicates the average pseudorapidity of the photon pair. (b)  Same as (a), but plotted using an expanded linear scale to illustrate the $\eta$ mass region. For the dashed line, the $\eta$ signal was removed from the simulation at the \textsc{pythia} level. (c) $A_N$ vs.\@ $M_{\gamma\gamma}$ for the above mass distribution. The error bars are statistical uncertainties only.}
\label{Mspect} 
\end{figure}

\begin{figure}[t]
\includegraphics[width=.48\textwidth]{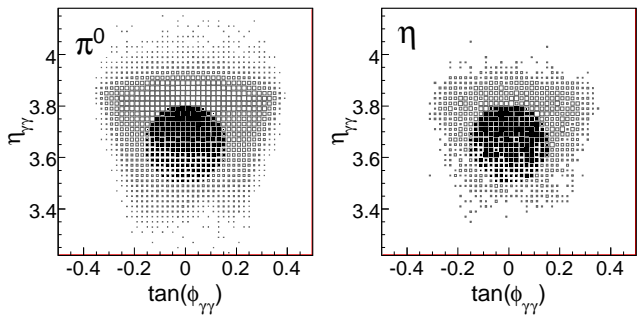}
\caption{Pseudorapidity vs. tangent of the azimuthal angle of the di-photon center of mass, for $E_{\gamma\gamma} > 50$ GeV. LEFT: $0.08 < M_{\gamma\gamma} < 0.19$ GeV/$c^2$, RIGHT: $0.45 < M_{\gamma\gamma} < 0.65$ GeV/$c^2$.  The filled boxes indicate events that pass the center cut (Eq.\@ (\ref{eq:center})). }
\label{etaphi} 
\end{figure}

The top two panels of Fig.\@ \ref{Mspect} show data-simulation comparisons of the di-photon invariant mass spectra. The ``center cut", so named because it covers roughly the central region of the FPD acceptance, is imposed on all event samples in order to enhance the $\eta$ meson acceptance relative to the background. It is defined as
\begin{equation}\label{eq:center}
(\eta_{\gamma\gamma}-3.65)^2 + \tan^2(\phi_{\gamma\gamma}) < 0.15 ,
\end{equation}
where $\eta_{\gamma\gamma}$ is the pseudorapidity of the di-photon center of mass relative to the polarized beam, and $\phi_{\gamma\gamma}$ is its azimuthal angle. The distributions of $\pi^0$ and $\eta$ events in the FPD, and the subset of each that pass the center cut are shown in Fig.\@ \ref{etaphi}. A full simulation based on \textsc{pythia} 6.222 and \textsc{geant} 3 was compared to the data. The reflectivity and absorption properties of the aluminized mylar wrapped glass blocks were varied to minimize the discrepancies between the photon shower shape in the simulation and that measured in the data. While detailed knowledge of the glass-mylar interface remains a limiting factor in the precise modeling of the shower development, the agreement in the widths of mass peaks between the simulation and data has been improved significantly over previous analyses \cite{Adams:2003fx, Adams:2006uz,:2008qb}. Furthermore, the data-simulation agreement in the continuum region between the $\pi^0$ and $\eta$ peaks is very good, allowing for a simulation-based background estimation for the $\eta$ signal. Corrections for the remaining data-simulation discrepancies in mass resolution were applied to the cross-section measurements. The $\eta$ to $\pi^0$ cross-section ratio in the simulation has been set at 0.45 to be consistent with the data. The bottom panel of Fig.\@ \ref{Mspect} shows the invariant mass dependence of $A_N$, which exhibits a suppression in the continuum region. Within the large statistical uncertainty, the asymmetry for this region does not show a significant $x_F$ dependence. In the simulation, this mass region is dominated by approximately equal contributions from a pair of photons from two different $\pi^0$ decays, and a charged hadron combined with a photon.

The energy resolution of the FPD is estimated to be about 7 to 8\% of the total energy based on the comparison of invariant mass and di-photon separation distributions between data and Cherenkov shower simulation. Coupled to the rapidly falling cross-section in energy, more than half of events in any measured energy bin originate from lower true energy bins. For the cross-section measurements, we unfolded the energy smearing by applying the Bayesian iterative method \cite{D'Agostini:1994zf} to the smearing matrices obtained from the simulation. The unfolding procedure combines the statistical and systematic uncertainties from the original data points.

\begin{figure}[t]
\includegraphics[width=.48\textwidth]{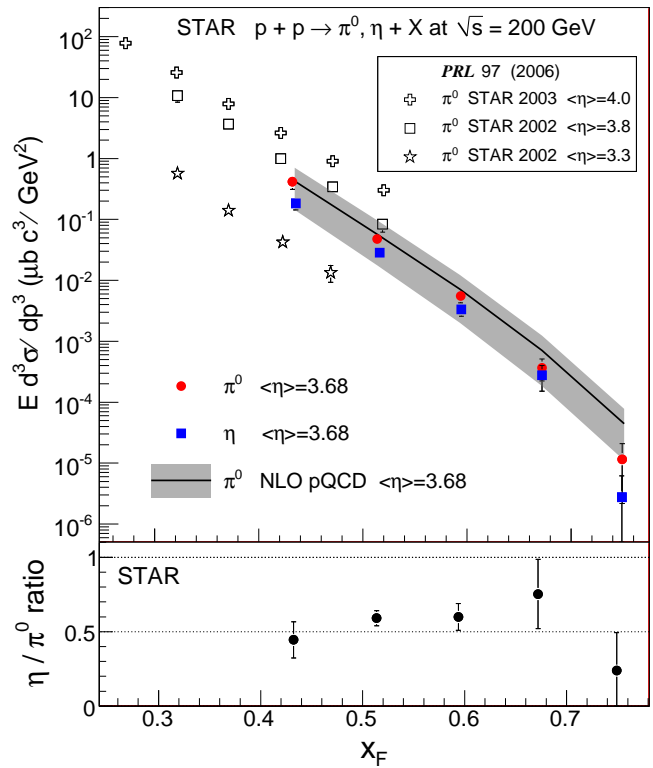}
\caption{(color online) Differential production cross-sections for $\pi^0$ and $\eta$ at average pseudorapidity of 3.68. Also shown are the previously published STAR results for similar kinematics \protect\cite{Adams:2006uz} and a NLO pQCD calculation of the $\pi^0$ cross-section \cite{vogel}. The error band represents the uncertainty in the calculation due to scale variations. The $\eta$ to $\pi^0$ cross-section ratio is shown in the bottom panel. The error bars indicate the total statistical and systematic uncertainties.}
\label{Xsec} 
\end{figure}

The upper panel of Fig.\@ \ref{Xsec} shows the differential cross-sections for $\pi^0$ and $\eta$. The center cut (Eq.\@ (\ref{eq:center})) was imposed on both mesons. Full \textsc{pythia} + \textsc{geant} simulations were used to obtain the detector efficiency corrections including the $\eta \rightarrow 2\gamma$ branching ratio. Also shown are the previously published STAR results for the $\pi^0$ cross-section in similar kinematic regions. The error band corresponds to the NLO pQCD theory prediction for the $\pi^0$ cross-section \cite{vogel}, based on the CTEQ6M5 parton distribution function \cite{Tung:2006tb} and the DSS fragmentation function \cite{deFlorian:2007aj}. The uncertainty for the theory prediction was obtained by increasing the factorization and renormalization scales from $\mu=p_T$ to $\mu=2p_T$. We note that the DSS fragmentation function includes in the fit the previously published STAR results at pseudorapidity of 3.3 and 3.8 \cite{Adams:2003fx}, along with other RHIC results. The error bars include both statistical and systematic uncertainties. The major sources of systematic uncertainties are the absolute energy calibration uncertainty of 3\%, which dominates the $\pi^0$ cross-section, and the uncertainty from the unfolding process, which dominates the $\eta$ cross-section at high energies. The normalization uncertainty was estimated at 12.5\%, including the uncertainty of the BBC coincidence cross-section of 7.6\% \cite{PhysRevLett.91.172302}.

The lower panel of Fig.\@ \ref{Xsec} shows the $\eta$ to $\pi^0$ cross-section ratio, which is found to be around 50\%. The error bars include both statistical and systematic uncertainties. The latter is dominated by the 1.5\% relative energy scale uncertainty, caused by the acceptance differences for $\pi^0$ and $\eta$ decay photons, and the localized variations in cell to cell calibration. The absolute calibration is common to both mesons, and largely cancels for the ratio. 

In pQCD, large-$x_F$ production of both $\pi^0$ and $\eta$ arises from hard-scattered partons fragmenting into mesons with large momentum fraction $z$ (ratio of hadron momentum to the momentum of its parent parton). The fragmentation process generally does not depend on the details of the hard scattering, and a single set of pion fragmentation functions explains a wide variety of RHIC data \cite{Adler:2003pi2, Adare:2007pi, Abelev:2010pi}. While there are currently no NLO pQCD predictions for the forward $\eta$ production cross-section, we note that our measurement of the $\pi^0$ cross-section is consistent with the NLO prediction, and the $\eta/\pi^0$ cross-section ratio is consistent with the recent NLO pQCD extraction of the $\eta$ fragmentation function \cite{Aidala:2010bn}.

\begin{figure}[t]
\includegraphics[width=.48\textwidth]{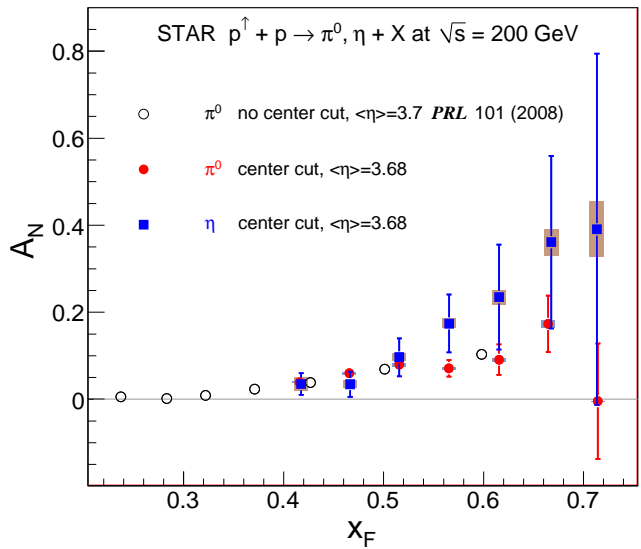}
\caption{(color online) $A_N$ vs.\@ $x_F$ at average pseudorapidity of 3.68 for $\pi^0$ and $\eta$. Also shown are the previously published results for $\pi^0$ at lower $x_F$, derived from the same data set but without the center cut \protect\cite{:2008qb}. The error bars are statistical uncertainties only. The error boxes indicate the systematic uncertainties. }
\label{AN} 
\end{figure}

Figure \ref{AN} shows the $A_N$, calculated using the ``cross ratio" formula \cite{Ohlsen197341,:2008qb}, as a function of $x_F$ for $\pi^0$ and $\eta$ after correcting for the underlying background. Also shown is the previous STAR result \cite{:2008qb} for $A_N$($\pi^0$) at lower $x_F$, which utilized the same data set as the current analysis but without the center cut. The two $\pi^0$ results are consistent within their correlated errors. The background correction, which only significantly affects the $\eta$ asymmetry at medium energy, is obtained from a simulation sample corrected for the $\eta$ yield and mass resolution, and the assumed background $A_N$ of $0.005 \pm 0.016$ extracted from Fig.\@ \ref{Mspect}(c). The error bars indicate statistical uncertainties only, while the error boxes indicate the systematic uncertainties. The main source of the systematic uncertainty is the background correction; the polarization uncertainty is negligible in comparison. The $A_N$ for negative $x_F$ is consistent with zero for both mesons.

The comparison of $A_N$ for $\pi^0$ and $\eta$ mesons is of particular interest given their similar up and down quark content, with wave-functions of both mesons containing $u\bar{u}$ and $d\bar{d}$ pairs. The $\eta$ differs from the $\pi^0$ mainly in that it is in an isospin singlet state, and that it contains $s\bar{s}$ in the wave function. The latter results in $\eta$ being significantly more massive than the $\pi^0$.  

In conclusion, STAR has measured the $x_F$ dependences of the cross-section and transverse single-spin asymmetries for $\pi^0$ and $\eta$ mesons produced at an average pseudorapidity of 3.68 in $\sqrt{s}=200$ GeV polarized proton collisions. For $0.55 < x_F < 0.75$, $A_N$($\eta$) ($0.210 \pm 0.056$) is 2.2 standard deviations larger than $A_N$($\pi^0$) ($0.081 \pm 0.016$). In this kinematic region, both the $\pi^0$ cross-section and the $\eta/\pi^0$ cross-section ratio are consistent with NLO pQCD expectations. This suggests that the measured $\eta$ asymmetry can be understood within the framework of pQCD. While several calculations exist for pion and kaon asymmetries \cite{Anselmino:1994tv,D'Alesio:2004up,Anselmino:2007fs,Anselmino:2008sga,Kouvaris:2006zy}, the first pQCD calculation of $A_N$ for the $\eta$ meson was performed only recently \cite{Kanazawa:2011bg}. This model generates an $\eta$ asymmetry that is substantially larger than that for the $\pi^0$ via a sizable initial-state twist-3 effect for strange quarks. The calculated $\eta$ asymmetry rises to about 12\% at $x_F$ of 0.4, well above our measured asymmetry, but then agrees quantitatively with the data for $x_F > 0.5$. It is yet unknown if a similar difference can arise from the fragmentation process via the Collins effect. A higher statistics measurement of the $A_N$ for the $\eta$ meson in this kinematic region is necessary to make a precise comparison to that for the $\pi^0$. Understanding the exact nature of these asymmetries can be further aided by complementary measurements of $A_N$ for final states that lack Collins contributions, such as jets and prompt photons. 

We thank the RHIC Operations Group and RCF at BNL, the NERSC Center at LBNL and the Open Science Grid consortium for providing resources and support. This work was supported in part by the Offices of NP and HEP within the U.S. DOE Office of Science, the U.S. NSF, the Sloan Foundation, a sponsored research grant from Renaissance Technologies Corporation, the DFG cluster of excellence `Origin and Structure of the Universe' of Germany, CNRS/IN2P3, FAPESP CNPq of Brazil, Ministry of Ed. and Sci. of the Russian Federation, NNSFC, CAS, MoST, and MoE of China, GA and MSMT of the Czech Republic, FOM and NWO of the Netherlands, DAE, DST, and CSIR of India, Polish Ministry of Sci. and Higher Ed., Korea Research Foundation, Ministry of Sci., Ed. and Sports of the Rep. of Croatia, and RosAtom of Russia.

\bibliographystyle{prsty}
\bibliography{etaAN_full_v4}

\begin{thebibliography}{10}

\bibitem{Kane:1978nd}
G.~L. Kane, J. Pumplin, and W. Repko, Phys. Rev. Lett. {\bf 41},  1689  (1978).

\bibitem{ZGS:1976}
R.~D. Klem {\it et~al.}, Phys. Rev. Lett. {\bf 36},  929  (1976).

\bibitem{AGS:1990}
S. Saroff {\it et~al.}, Phys. Rev. Lett. {\bf 64},  995  (1990).

\bibitem{Adams:1991rw}
D.~L. Adams {\it et~al.}, Phys. Lett. B {\bf 261},  201  (1991).

\bibitem{Adams:1991cs}
D.~L. Adams {\it et~al.}, Phys. Lett. B {\bf 264},  462  (1991).

\bibitem{Allgower:2002qi}
C.~E. Allgower {\it et~al.}, Phys. Rev. D {\bf 65},  092008  (2002).

\bibitem{Sivers:1989cc}
D.~W. Sivers, Phys. Rev. D {\bf 41},  83  (1990).

\bibitem{Collins:1992kk}
J.~C. Collins, Nucl. Phys. B {\bf 396},  161  (1993).

\bibitem{Qiu:1991pp}
J.-W. Qiu and G.~F. Sterman, Phys. Rev. Lett. {\bf 67},  2264  (1991).

\bibitem{Kouvaris:2006zy}
C. Kouvaris, J.-W. Qiu, W. Vogelsang, and F. Yuan, Phys. Rev. D {\bf 74},
  114013  (2006).

\bibitem{Kang:2011hk}
Z.-B. Kang, J.-W. Qiu, W. Vogelsang, and F. Yuan, Phys. Rev. D {\bf 83},
  094001  (2011).

\bibitem{:2008qb}
B.~I. Abelev {\it et~al.}, Phys. Rev. Lett. {\bf 101},  222001  (2008).

\bibitem{Adler:2005in}
S.~S. Adler {\it et~al.}, Phys. Rev. Lett. {\bf 95},  202001  (2005).

\bibitem{:2008mi}
I. Arsene {\it et~al.}, Phys. Rev. Lett. {\bf 101},  042001  (2008).

\bibitem{Airapetian:2004tw}
A. Airapetian {\it et~al.}, Phys. Rev. Lett. {\bf 94},  012002  (2005).

\bibitem{:2008dn}
M. Alekseev {\it et~al.}, Phys. Lett. B {\bf 673},  127  (2009).

\bibitem{Airapetian:2009hr}
A. Airapetian {\it et~al.}, Phys. Rev. Lett. {\bf 103},  152002  (2009).

\bibitem{Alekseev:2010cp}
M. Alekseev {\it et~al.}, Phys. Lett. B {\bf 692},  240  (2010).

\bibitem{Adams:1997dp}
D.~L. Adams {\it et~al.}, Nucl. Phys. B {\bf 510},  3  (1998).

\bibitem{Adams:2003fx}
J. Adams {\it et~al.}, Phys. Rev. Lett. {\bf 92},  171801  (2004).

\bibitem{Adams:2006uz}
J. Adams {\it et~al.}, Phys. Rev. Lett. {\bf 97},  152302  (2006).

\bibitem{Adare:2007pi}
A. Adare {\it et~al.}, Phys. Rev. D {\bf 76},  051106  (2007).

\bibitem{Adare:2008pi}
A. Adare {\it et~al.}, Phys. Rev. D {\bf 79},  012003  (2009).

\bibitem{Abelev:2009pi}
B.~I. Abelev {\it et~al.}, Phys. Rev. D {\bf 80},  111108  (2009).

\bibitem{Abelev:2010pi}
B.~I. Abelev {\it et~al.}, Phys. Rev. C {\bf 81},  064904  (2010).

\bibitem{Adler:2006bv}
S.~S. Adler {\it et~al.}, Phys. Rev. C {\bf 75},  024909  (2007).

\bibitem{Adare:2011eta}
A. Adare {\it et~al.}, Phys. Rev. D {\bf 83},  032001  (2011).

\bibitem{Arsene:2007jd}
I. Arsene {\it et~al.}, Phys. Rev. Lett. {\bf 98},  252001  (2007).

\bibitem{PhysRevLett.97.252001}
B.~I. Abelev {\it et~al.}, Phys. Rev. Lett. {\bf 97},  252001  (2006).

\bibitem{PhysRevLett.91.172302}
J. Adams {\it et~al.}, Phys. Rev. Lett. {\bf 91},  172302  (2003).

\bibitem{D'Agostini:1994zf}
G. D'Agostini, Nucl. Instrum. Meth. A {\bf 362},  487  (1995).

\bibitem{vogel}
W. Vogelsang, Private Communication  (2011).

\bibitem{Tung:2006tb}
W.~K. Tung {\it et~al.}, J. High Energy Phys. {\bf 02},  053  (2007).

\bibitem{deFlorian:2007aj}
D. de~Florian, R. Sassot, and M. Stratmann, Phys. Rev. D {\bf 75},  114010
  (2007).

\bibitem{Adler:2003pi2}
S.~S. Adler {\it et~al.}, Phys. Rev. Lett. {\bf 91},  241803  (2003).

\bibitem{Aidala:2010bn}
C.~A. Aidala {\it et~al.}, Phys. Rev. D {\bf 83},  034002  (2011).

\bibitem{Ohlsen197341}
G.~G. Ohlsen and P.~W. {Keaton Jr.}, Nucl. Instrum. Meth. {\bf 109},  41
  (1973).

\bibitem{Anselmino:1994tv}
M. Anselmino, M. Boglione, and F. Murgia, Phys. Lett. B {\bf 362},  164
  (1995).

\bibitem{D'Alesio:2004up}
U. D'Alesio and F. Murgia, Phys. Rev. D {\bf 70},  074009  (2004).

\bibitem{Anselmino:2007fs}
M. Anselmino {\it et~al.}, Phys. Rev. D {\bf 75},  054032  (2007).

\bibitem{Anselmino:2008sga}
M. Anselmino {\it et~al.}, Eur. Phys. J. A {\bf 39},  89  (2009).

\bibitem{Kanazawa:2011bg}
K. Kanazawa and Y. Koike, Phys. Rev. D {\bf 83},  114024  (2011).

\end{thebibliography}

\end{document}